\documentclass[5p,twocolumn]{elsarticle}

\usepackage{hyperref}
\usepackage{amsmath}

\journal{Physics Letters B}

\newcommand{\qweak}{$Q_{\rm weak}$}

\newcommand{\moller}{M\o ller}

\newcommand{\muA}{$\mu$A}

\begin{document}

\begin{frontmatter}

\title{A novel comparison of M\o ller and Compton electron-beam polarimeters}

\author[1]{J.~A.~Magee}
\author[2]{A.~Narayan}
\author[3]{D.~Jones}
\author[4]{R.~Beminiwattha}
\author[1]{J.~C.~Cornejo}
\author[3,5]{M.~M.~Dalton}
\author[1]{W.~Deconinck}
\author[2]{D.~Dutta}
\author[5]{D.~Gaskell\corref{a}}
\cortext[a]{Corresponding author}
\ead{gaskelld@jlab.org}
\author[6]{J.~W. Martin} 
\author[3]{K.~D.~Paschke}
\author[6,7]{V.~Tvaskis}
\author[8]{A.~Asaturyan}
\author[5]{J. Benesch}
\author[3]{G.~Cates}
\author[9]{B.~S.~Cavness}
\author[5]{L.~.A. ~Dillon-Townes}
\author[5]{G.~Hays}
\author[1]{J.~Hoskins}
\author[10]{E.~Ihloff}
\author[11]{R.~Jones}
\author[4]{P.~M.~King}
\author[12]{S.~Kowalski}
\author[13]{L.~Kurchaninov} 
\author[13]{L.~Lee}
\author[14]{A.~McCreary}
\author[6]{M.~McDonald}
\author[6]{A.~Micherdzinska}
\author[8]{A.~Mkrtchyan}
\author[8]{H.~Mkrtchyan}
\author[3]{V.~Nelyubin}
\author[7]{S.~Page}
\author[13]{W.~D.~Ramsay}
\author[5]{P.~Solvignon\fnref{b}}
\fntext[b]{Deceased}
\author[6]{D.~Storey}
\author[3]{W.~A.~Tobias}
\author[15]{E.~Urban}
\author[10]{C.~Vidal}
\author[4]{B. Waidyawansa}
\author[7]{P.~Wang}
\author[8]{S.~Zhamkotchyan}

\address[1]{College of William and Mary, Williamsburg, VA 23187}
\address[2]{Mississippi State University, Mississippi State, MS 39762, USA}
\address[3]{University of Virginia, Charlottesville, VA 22904, USA} 
\address[4]{Ohio University, Athens, OH 45701}
\address[5]{Thomas Jefferson National Accelerator Facility, Newport News, VA 23606, USA}
\address[6]{University of Winnipeg, Winnipeg, MB R3B2E9, Canada}
\address[7]{University of Manitoba, Winnipeg,  MB R3E0W3, Canada}
\address[8]{Yerevan Physics Institute, Yerevan, 375036, Armenia}
\address[9]{Angelo State University, San Angelo, TX 76903, USA}
\address[10]{MIT Bates Linear Accelerator Center, Middleton, MA 01949, USA}
\address[11]{University of Connecticut, Storrs, CT 06269, USA}
\address[12]{Massachusetts Institute of Technology, Cambridge, MA 02139, USA}
\address[13]{TRIUMF, Vancouver, BC V6T~2A3, Canada}
\address[14]{University of Pittsburgh, Pittsburgh, PA 15260, USA}
\address[15]{Hendrix College, Conway, AR 72032, USA}

\begin{abstract}
We have performed a novel comparison between electron-beam polarimeters based on
M\o ller and Compton scattering. A sequence of electron-beam polarization measurements
were performed at low beam currents ($<$~5~$\mu$A) during the \qweak{} experiment
in Hall C at Jefferson Lab.  These low current measurements
were bracketed by the regular high current (180~$\mu$A) operation of the Compton
polarimeter. All measurements were found to be consistent within experimental
uncertainties of 1\% or less, demonstrating that electron polarization does not
depend significantly on the beam current.  This result lends confidence to the
common practice of applying M\o ller measurements made at low beam currents
to physics experiments performed at higher beam currents. The agreement between two
polarimetry techniques based on independent physical processes sets an important
benchmark for future precision asymmetry measurements that require sub-1\%
precision in polarimetry.
\end{abstract}


\end{frontmatter}

\section{Introduction}
\label{sec:intro}
Polarized electrons have become an essential tool in nuclear and particle
physics experiments seeking to understand the fundamental forces of nature. They
are used to address a wide variety of topics ranging from the internal structure
of nucleons to precision tests of the Standard Model.  Typically these studies
need longitudinal beam polarization, where the electron spin is orientated
parallel or anti-parallel to the beam momentum.  Over the last several decades
technological progress has resulted in dramatic improvement in the precision of
experiments that utilize polarized beams.  Knowledge of the beam polarization is
an important source of experimental uncertainty in these experiments. Future
measurements~\cite{MOLLER,SOLID,P2} will require 0.5\% precision in the beam
polarization. Advances in electron-beam polarimetry are a key driver for the future
improvements in precision spin physics.

All techniques used to determine the electron-beam polarization at accelerator
energies exploit the spin dependence  of a scattering process by measuring 
the difference in the electron scattering rate for two possible helicity
configurations.  The ratio of this difference to the sum of the scattering rates is
called the asymmetry, and is directly proportional to the beam polarization and the
analyzing power of the scattering process. Hence, for scattering processes whose
analyzing power can be precisely calculated, we can determine the polarization of the
beam by measuring the asymmetry.

It is desirable that the spin-dependent scattering process have a large rate and a
slowly varying analyzing power. When polarized targets are used, they should be stable
and highly polarized with easily measurable polarization. It is also desirable that the
polarimetry technique be non-invasive to the physics experiment and use the same beam,
i.e. the same energy, current and location, as the physics experiment.
The two processes most commonly employed in electron-beam polarimetry are M\o ller
scattering, from spin-polarized electrons in magnetic materials, and Compton scattering,
from circularly polarized laser photons.

None of the readily available polarimetry techniques have all of the desirable
properties. For example, polarimeters based on M\o ller scattering can be
operated only at low currents making this technique invasive to the experiment,
while the analyzing power in Compton scattering varies rapidly (even changing sign)
as a function of the energy of the scattered particles.
Thus, in order to achieve the desired high accuracy, multiple independent and
high precision polarimeters have to be used in concert.  The most recent
experiment to employ both M\o ller and Compton polarimeters was the \qweak{} experiment,
a parity--violating electron scattering experiment in Hall-C at Jefferson Lab
(JLab)~\cite{qweakprl,qweaknim,compton2}. The \qweak{} experiment aims to test the Standard
Model of particle physics by providing a first precision measurement of the weak vector
charge of the proton, from which the weak mixing angle will be extracted with the
highest precision to date away from the Z$^0$ pole. Knowledge of the electron-beam
polarization is one of the largest experimental corrections for the
\qweak{} experiment. To achieve the desired precision the experiment used an existing
high-precision M\o ller polarimeter and a new Compton polarimeter to continuously
monitor the electron-beam polarization. The M\o ller polarimeter was used
intermittently throughout the experiment, operating at a beam current of a few
microamps, while the Compton polarimeter monitored the beam polarization at the exact
running conditions of the \qweak{} experiment, which included beam currents of up to
$\sim180$~\muA, at a beam energy of 1.16~GeV. 

The two polarimeters have very different analyzing powers and systematic uncertainties.
A precise comparison of the polarization measured by the M\o ller and Compton,
in quick succession and at the same beam currents provides an essential
cross-check between the two techniques.
Further, comparison of the measured beam polarization at different beam
currents can be used to verify the often used assumption that the electron-beam
polarization is independent of beam current. In this letter we report the results
from a series of measurements where M\o ller, followed by Compton, followed by M\o ller
measurements were performed at identical low beam currents in rapid succession. In
addition the sequence of measurements was bracketed by Compton measurements at
high currents.

A previous comparison of multiple electron-beam polarimeters at Jefferson Lab has been reported
in Ref.~\cite{grames}. In that study, measurements were made using a Mott polarimeter in the
accelerator injector, M\o ller polarimeters in Halls A, B, and C, and a Compton polarimeter in
Hall A (the Hall-C Compton polarimeter did not exist at that time). While the polarizations
extracted by the various devices were deemed compatible within their systematic uncertainties,
only the Hall-C M\o ller polarimeter was capable of making measurements with a systematic
uncertainty better than 1\%. In addition, the measurements using the M\o ller polarimeters
were performed at different beam conditions than the Compton polarimeter.

The measurement described in this letter is the first comparison of two polarimeters
capable of sub-1\% systematic errors, performed in the same experimental hall under
identical beam conditions. The entire sequence of measurements allows us to compare
the Compton and M\o ller measurements at low currents, as well as to compare them to the
Compton measurements at high current.

\section{The Hall-C M\o ller Polarimeter}
\label{sec:mol}

\begin{figure}[hbtp!]
{\includegraphics*[width=8.5cm]{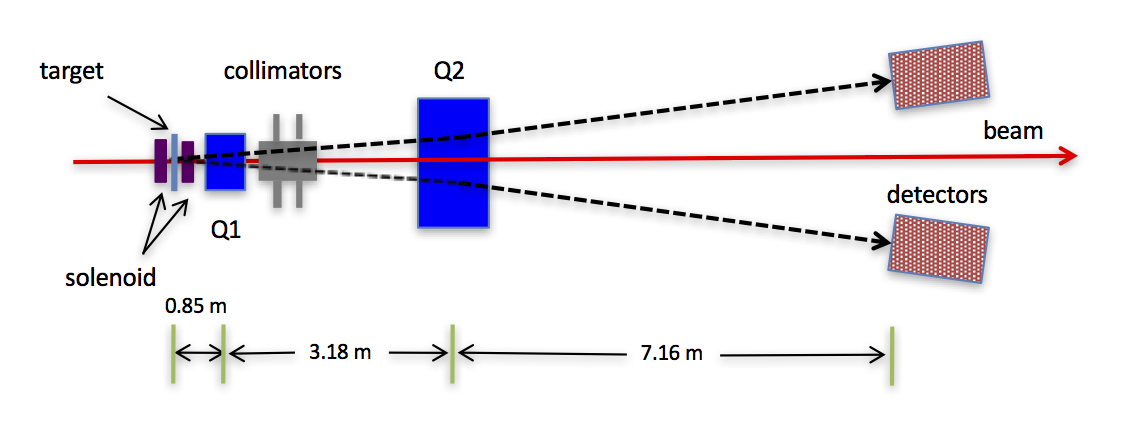}}
\caption[]{Schematic of the JLab Hall-C M\o ller polarimeter.  Note that the quadrupole positions
are slightly different than described in Ref.~\cite{moller1}.}
\label{fig:fig1}
\end{figure}

The Hall-C M\o ller polarimeter is designed to provide an absolute polarization
measurement with a statistical precision of better than 0.5\% within a few minutes.
Fig.~\ref{fig:fig1} depicts a schematic of the polarimeter.  A 3.5~T superconducting
solenoid magnet is used to polarize a 1--4 $\mu$m thick, pure iron foil out-of-plane with
an applied magnetic field well above the 2.2~T required for magnetic saturation of pure iron.
The maximum of the analyzing power for M\o ller scattering occurs at a scattering angle
of $90^{\circ}$ in the center-of-mass frame and the polarimeter is designed to optimize
the acceptance for these kinematics.  A pair of quadrupole magnets focus the scattered
and recoiling atomic electrons onto the detectors.  Detection of both electrons in
coincidence reduces backgrounds due to Mott scattering.  A set of movable collimators
placed between the quadrupole magnets is employed to further reduce the Mott background
without affecting the acceptance of M\o ller electrons.  The effective analyzing power for the
Hall-C M\o ller polarimeter is $\approx0.064$ at 1.16~GeV, with some variation (typically less
than 0.5\%) due to beam position variation. The details of the JLab Hall-C M{\o}ller
polarimeter are described in Ref.~\cite{moller1,moller2}

\begin{figure*}[htbp]
\includegraphics*[height=17cm,angle=-90]{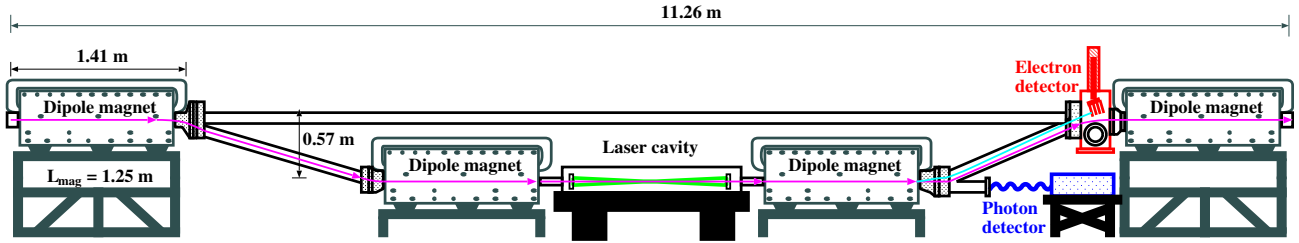}
\caption[]{Schematic of the JLab Hall-C Compton polarimeter. Figure taken
  from~\cite{compton2}.} 
\label{fig:fig2}
\end{figure*}

\section{The Hall-C Compton Polarimeter}
\label{sec:cpol}
A schematic of the Compton polarimeter in Hall C at JLab is shown in
Fig.~\ref{fig:fig2}. The JLab Hall-C Compton polarimeter was designed to continuously
monitor the beam polarization at high beam currents with better than 1\% statistical
uncertainty per hour. It consists of four identical dipole magnets forming a magnetic
chicane that displaces a 1.16~GeV electron beam vertically downward  by 57~cm. A high
intensity ($\sim$ 1--2~kW) beam of $\sim$ 100\% circularly polarized photons is provided
by an external low-gain Fabry-P\'{e}rot laser cavity which consists of an 85~cm long
optical cavity with a gain between 100 and 200, coupled to a green (532~nm), continuous
wave, 10~W laser (Coherent VERDI). The laser light is focused at the interaction region
($\sigma_{\text{waist}}\sim$~90~$\mu$m), where it is larger than the electron beam envelope
($\sigma_{x/y} \sim$~40~$\mu$m).  The laser is operated in 90 second cycles, where it is
active for 60 seconds (laser on period) and blocked off (laser off period) for the rest
of the cycle. The laser off data are used to measure the background.

At the electron beam energy of 1.16 GeV used for these studies, the maximum scattered photon
energy is approximately 46 MeV, while the maximum
separation between the primary electron beam and the Compton scattered electrons,
just upstream of the fourth dipole, is $\sim$ 17 mm.  The displacement of the scattered
electron with respect to the primary electron beam is detected by a set of four diamond
micro-strip detectors. The proximity of the detectors to the primary beam allows them
to capture most of the energy spectrum of the scattered electrons. The data analysis
technique exploits track finding, the high granularity of the electron detector and its
large acceptance to fit the shape of the measured asymmetry spectrum to the precisely
calculable Compton asymmetry.   

A calorimeter consisting of a 2$\times$2 matrix of 3~cm$\times$3~cm$\times$20~cm
PbWO$_4$ scintillating crystals attached to a single photo-multiplier tube was used to
measure the scattered photon energy. The signal from the photon detector is digitally
integrated with no thresholds over a full helicity state ($\sim$ 1 ms) using a 200~MHz
flash analog to digital converter.

Details on the Compton polarimeter can be found in
Ref.~\cite{qweaknim, compton2, compton1, compton1b,compton1c}.  Since the
electron detector provided the highest precision and most reliable measurements
of the beam polarization from the Compton polarimeter, the results discussed here
are from the electron detector only.

\section{Data Analysis and Results}
\label{sec:analysis}
The electron beam helicity was reversed at a rate of 960 Hz in a pseudo-random
sequence, using an electro-optic Pockels cell in the laser optics of the polarized
electron photoemission gun~\cite{source1, source2}. In addition, a half-wave plate
in the polarized source laser optics was inserted or removed about every 8 hours to
reverse the beam helicity as a systematic check.  These reversals change the sign
of the beam polarization, but the magnitude of the polarization was found to be
independent of the half-wave plate.  The analysis
procedure to extract beam polarization from the \moller{} measurements is described in
Ref.~\cite{moller1,moller2} while the procedure for the Compton measurements is  described in
Ref.~\cite{compton2}.

Typically, \moller{} measurements are conducted at low beam currents ($<2$~\muA)
to minimize the foil depolarization due to beam heating, whereas
Compton polarimeters achieve their best statistical precision when operated at the
maximum beam current.  Comparing the two
under identical conditions  required finding a suitable ``compromise'' current. At this
beam current, target foil depolarization effects due to beam heating would be minimal
for the \moller{} while still enabling the Compton to achieve adequate statistics in
a reasonable time. This study consisted of two \moller{} current scans, with an 8-hour
series of Compton measurements in-between.

The temperature increase in the M\o ller target foil due to the power deposited by the
electron beam was  determined by solving the 1D-radial heat equation
using the known (temperature dependent) value of the thermal conductivity of iron and
knowledge of the approximate size of the electron beam. The calculation assumed that
radiative cooling effects were small and that the primary cooling mechanism was via
conduction through the target holder.

After calculating the target's temperature rise, the target depolarization was
calculated using empirical fits which were shown to agree well with previously
published measurements of the temperature dependence of the magnetization of
iron~\cite{iron_mag}.

At beam currents of 1~$\mu$A or smaller, the foil depolarization due to beam heating
was estimated to be less than 0.14\%.  At the highest currents used in this study, the
depolarization grows to almost 1\%.

\begin{figure}[hbtp!]
{\includegraphics*[width=8.5cm]{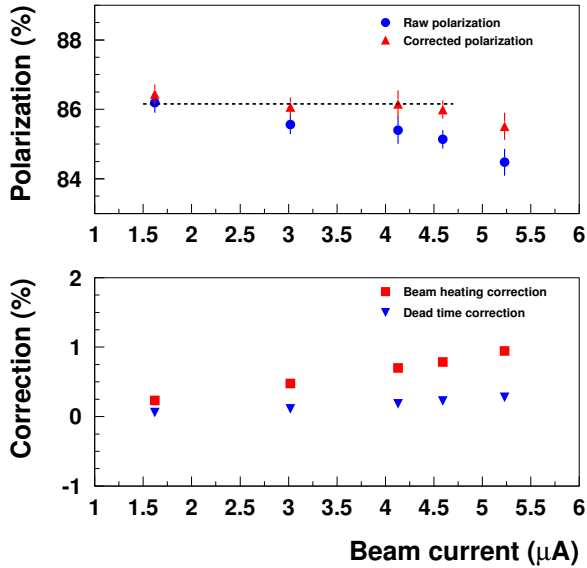}}
\caption[]{Top: Electron-beam polarization as determined by the \moller{}
  polarimeter as a function of current. The blue circles are the raw measurements,
  while the red triangles include corrections for dead time and target heating effects.
  Error bars denote statistical uncertainties only.
  The dashed line is the best fit of the corrected data for all points at currents less
  than 5~\muA.  The average is 86.16$\pm$0.15\% (stat), with a $\chi^2/dof~$ of 0.5
  (associated probability 0.69). 
  \\ Bottom:
  Size of correction vs. current, in percent. Red squares are the beam heating
  correction, while blue inverted triangles are the dead time correction.}
\label{fig:moller_heating}
\end{figure}

Dead time is the only other effect that depends significantly on beam current.
Corrections for target heating and dead-time, and their effect on the measured
polarization, are shown in Fig.~\ref{fig:moller_heating} (note that only
statistical uncertainties are shown in the top panel of Fig.~\ref{fig:moller_heating}).
The uncertainty on the target heating correction is estimated to be about 30\% of the size
of the correction.  At 4.5~\muA, the maximum current used for the M\o ller and Compton comparison,
the impact of the target heating uncertainty is $\Delta P/P$ = 0.24\% - this value is used
when comparing the M\o ller and Compton results.

The largest systematic uncertainty for the \moller{} measurement comes from the effect
of atomic Fermi motion of the electrons (the Levchuk effect) \cite{levchuk}. Other
significant uncertainties come from beam position and angle on target, which were
determined from special systematic studies of the analyzing power dependence on beam
position. These are discussed in \cite{magee}.

The full list of systematic uncertainties for the \moller{} polarimeter during this
study is given in Table \ref{tab:moller}. A discussion of the typical conditions is
given in \cite{qweaknim}. The major difference is the absence of a rather large
uncertainty of 0.50\% for extrapolating to higher currents. Since both polarimeters
ran at approximately the same current ($<5$~\muA) for this cross-calibration, a
high-current extrapolation was not necessary.

\setlength{\tabcolsep}{4pt}
\begin{table}[htb!]
\caption{Systematic Uncertainties of the M\o ller Polarimeter}
\label{tab:moller}
\begin{center}
\begin{tabular}{|l|c|c|}
\hline
Source & Uncertainty & $\Delta$P/P\% \\\hline
      Beam position X            & 0.2 mm       & 0.14          \\ 
      Beam position Y            & 0.2 mm       & 0.28          \\ 
      Beam angle X               & 0.5 mrad     & 0.10          \\ 
      Beam angle Y               & 0.5 mrad     & 0.10          \\ 
      Q1 current                 & 2\%          & 0.07          \\ 
      Q3 current                 & 2\%          & 0.05          \\ 
      Q3 position                & 1 mm         & 0.10          \\ 
      Multiple scattering        & 10\%         & 0.01          \\ 
      Levchuk effect             & 10\%         & 0.33          \\ 
      Fixed collimator positions & 0.5 mm       & 0.03          \\ 
      Beam heating of target     & 30\%         & 0.24          \\ 
      B-field direction          & 2 degrees    & 0.14          \\ 
      B-field strength           & 5\%          & 0.03          \\ 
      Spin polarization in Fe    & -----        & 0.25          \\ 
      Electronic D.T.            & 100\%        & 0.045         \\ 
      Solenoid focusing          & 100\%        & 0.21          \\ 
      Solenoid position (x,y)    & 0.5 mm       & 0.23          \\ 
      Monte Carlo statistics     & -----        & 0.14          \\ \hline
      \textbf{Total}             & ~            & \textbf{0.71} \\ \hline
\end{tabular}
\end{center}
\end{table}

For the Compton measurements, as described in Ref.~\cite{compton1}, the yield
asymmetry measured by the electron detector for each $\sim$ 1 hour long interval (run)
was compared to the theoretical Compton asymmetry for each detector strip, with the
beam polarization ($P_e$) and the non-integer strip number corresponding to the maximum
displaced electrons ($n_{CE}$), or the Compton edge, as the two independent parameters.

\begin{figure}[hbtp!]
{\includegraphics*[width=8.0cm]{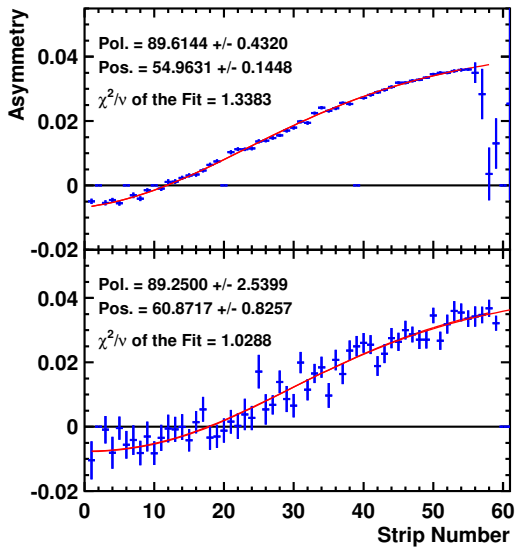}}
\caption[]{The typical Compton asymmetry (blue) and fit (red) for high current (upper)
  and low current (lower). The extracted polarization, location of the Compton edge
  and the $\chi^2$ per degree of freedom for the fit are also indicated in the figure.} 
\label{fig:fig3}
\end{figure}

The typical asymmetry obtained from the electron detector fit to the calculated
asymmetry is shown in Fig.~\ref{fig:fig3}. The upper panel shows the asymmetry and fit
for high current, while the bottom panel shows the asymmetry and fit at low current.  

DAQ inefficiency, from such things as dead-time and the trigger-forming algorithm, was
determined using a simulation of the DAQ system~\cite{compton2}.  The DAQ simulation
was used to determine a correction factor for the detector yield based on the aggregate
detector rate. The corrections were negligible for the low current data and $<1$\% for
the high current data. An independent analysis of a separate data stream, using only raw
hits in the electron detector rather than triggers that met the tracking definitions,
was used to validate these results over a wide range of rates. The two analyses,
in principle, have very different rate sensitivities, and their comparison was used to
demonstrate the validity of these rate-dependent corrections. The DAQ-related systematic
uncertainties are listed in Table.~\ref{tab:tab2}.
     
A full Monte Carlo simulation of the Compton polarimeter using the GEANT3~\cite{g3}
detector simulation package was used to validate the analysis procedure and to study a
variety of sources of systematic uncertainties~\cite{compton1}.  The list of
contributions is shown in Table ~\ref{tab:tab2}. From these simulation studies we have
determined that the net systematic uncertainty of the extracted beam polarization is
0.59\%~\cite{compton2}. 

\setlength{\tabcolsep}{4pt}
\begin{table}[htb!]
\caption{Systematic Uncertainties of the Compton Polarimeter}
\label{tab:tab2}
\begin{center}
\begin{tabular}{|l|c|c|}
\hline
Source & Uncertainty & $\Delta$P/P\% \\\hline
Laser Polarization              & 0.18\%              &       0.18    \\
Magnetic field                  & 0.0011 T            &       0.13    \\
Beam energy                     & 1 MeV               &       0.08    \\
Detector z position             & 1 mm                &       0.03    \\
Trigger multiplicity            & 1-3 plane           &       0.19    \\
Trigger clustering              & 1-8 strips          &       0.01    \\
Detector tilt(w.r.t x, y, z)    & 1 degree            &       0.06    \\
Detector efficiency             & 0.0 - 1.0           &       0.1     \\
Detector noise                  & up to 20\% of rate  &       0.1     \\
Fringe field                    & 100\%               &       0.05    \\
Radiative corrections           & 20\%                &       0.05    \\
DAQ inefficiency correction     & 40\%                &       0.3     \\
DAQ inefficiency pt.-to-pt.     & -----               &       0.3     \\
Beam vert. pos. variation       & 0.5 mrad            &       0.2     \\
Helicity correl. beam pos.      & 5 nm                &       $<$ 0.05\\
Helicity correl. beam angle     & 3 nrad              &       $<$ 0.05\\
Spin precession in chicane      & 20 mrad             &       $<$ 0.03\\ \hline
\textbf{Total}                  &                     &    \textbf{0.59} \\\hline
\end{tabular}
\end{center}
\end{table}

\begin{figure}[hbtp!]
{\includegraphics*[width=8.5cm]{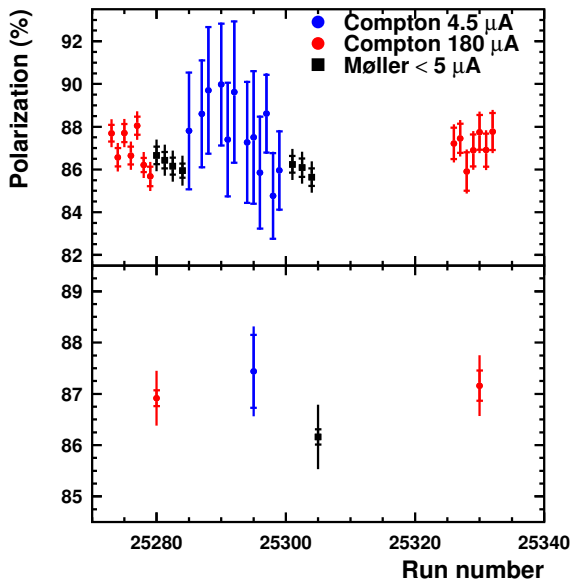}}
\caption[]{(top) Polarization measured at 4.5~$\mu$A along with neighboring high
  current runs and the M\o ller measurements taken at beam currents less than 5~\muA.
  (bottom) The average of high current and low current Compton and the neighboring
  M\o ller measurements.}
\label{fig:mcmfinal}
\end{figure}

At low beam currents, the measurement in the Compton polarimeter is hampered by poor
statistics (as seen in Fig.~\ref{fig:fig3}). On the other hand, increasing the beam
current results in larger systematic uncertainties in the M\o ller measurement, due to
target heating. As mentioned earlier from the current scan of the \moller{} polarimeter,
it was determined that the highest beam current where the heating effects do not
dominate is $\sim$ 4.5~$\mu A$. Therefore, the M\o ller - Compton - M\o ller (MCM)
sequence of measurements described here, were performed at a beam current of
4.5~$\mu$A. In order to keep the electron-beam optics identical for the M\o ller and
Compton measurements, the beam was transported through the Compton chicane for both
polarimeters during the MCM comparison. The beam transport parameters were identical
during the comparison measurements. A slow position feedback loop, which locked the
electron-beam position at the laser interaction point, was used for the Compton
measurements but disabled for the M\o ller measurements.
Fig.~\ref{fig:mcmfinal} shows the polarization
extracted from the consecutive M\o ller, Compton and M\o ller measurements using the
same low current ($\sim$ 4.5~\muA{} for the Compton measurements and $
\sim$ 1.5-4.5~\muA{} for the M\o ller measurements) beam. The results demonstrate
that the polarization measured by the M\o ller and Compton polarimeters are consistent
within experimental uncertainties (about 1\% relative for the Compton and 0.73\% for
the M\o ller) at low beam current. Also plotted in Fig.~\ref{fig:mcmfinal} are the two
adjacent high current ($\sim$ 180~$\mu$A) Compton measurements, during which we do not
expect the electron-beam polarization to have changed. The consistency between the low
current and high current measurements indicate that within experimental uncertainties
the beam polarization does not vary with beam current. The polarization obtained from
the Compton measurements for both high current runs and low current runs were averaged
separately and are shown in the lower panel of Fig.~\ref{fig:mcmfinal} and listed in
Table~\ref{tab:mcmResult}. In these plots the inner error bar shows the statistical
uncertainty while the outer error bar shows the total uncertainty given by the
quadrature sum of statistical and systematic uncertainties. A detailed analysis of the
Compton and M\o ller data, properly accounting for correlated systematic uncertainties
for the Compton measurements, indicates a 1-$\sigma$ upper limit of 0.98\% (relative)
for the change in beam polarization between 4.5 and 180~$\mu$A. This result does not rule
out some small dependence of electron-beam polarization on current, but does demonstrate
that polarization measurements made at low currents can safely be applied at higher
currents without a significant increase in systematic uncertainty.

\setlength{\tabcolsep}{4pt}
\begin{table}[htb!]
  \caption{Mean polarization measurements during cross calibration. The first M\o ller
    entry is averaged over all beam currents less than 5~\muA, while the second is at
    a fixed beam current of 4.5~\muA.}
\label{tab:mcmResult}
\begin{center}
\begin{tabular}{|l|c|c|c|c|}
\hline
Type            & Current & Mean      & Stat.   & Total   \\
                & $\mu$A  & pol. (\%) & uncert. & uncert. \\
\hline
Compton         & 180     & 86.92     & 0.15    & 0.53    \\
M\o ller (avg.) & 3.3     & 86.16     & 0.15    & 0.63    \\
M\o ller        & 4.5     & 86.00     & 0.27    & 0.67    \\
Compton         & 4.5     & 87.44     & 0.71    & 0.88    \\
Compton         & 180     & 87.16     & 0.29    & 0.59    \\
\hline
\end{tabular}
\end{center}
\end{table}

\section{Conclusions}
\label{sec:res}
We have compared the polarization obtained from consecutive measurements using a
M\o ller and a Compton polarimeter at low beam currents.  These low current measurements
were bracketed by the regular high current operation of the Compton polarimeter.
All measurements were found to be consistent within experimental uncertainties,
demonstrating that the electron-beam polarization does not depend on beam current to
better than 1\% for a beam current range of 175~\muA{}. These results give confidence
in the use of M\o ller measurements made at low beam currents for physics experiments
carried out at much higher current.  In addition, the demonstration of the consistency
of two high-precision electron-beam polarimeters under identical beam conditions is a
significant step forward in verifying the accuracy of electron-beam polarimetry at the
level required by future high-precision measurements of spin-dependent asymmetries, such
as the SOLID and MOLLER experiments at Jefferson Lab or the P2 experiment at Mainz.

\section{Acknowledgments}
We thank the \qweak{} collaboration for their support and cooperation in the execution
of the tests described in this paper.

This work was funded by the U.S. Department of Energy, including contract 
\# AC05-06OR23177, under which Jefferson Science Associates, LLC operates Thomas
Jefferson National Accelerator Facility, and by the
Natural Sciences and Engineering Research Council of Canada (NSERC).  We wish to thank
the staff of JLab, TRIUMF, and Bates, for their vital support.

\section*{References}

\bibliography{mcm_comparison}

\end{document}